\documentclass{JHEP3}
\usepackage{amsmath}
\usepackage{amssymb}

\usepackage{amsmath,amssymb,graphics}
\usepackage{epsfig,multicol}
\usepackage{graphicx}
\usepackage{epsfig,subfigure}

\title{de Sitter Thick Brane Solution in Weyl Geometry}

\author{Yu-Xiao Liu,
        Ke Yang\footnote{Corresponding author.},
        Yuan Zhong
        \\
        Institute of Theoretical Physics,
        Lanzhou University, Lanzhou 730000,
        People's Republic of China\\
  E-mail: \email{liuyx@lzu.edu.cn},
          \email{yangke09@lzu.cn},
          \email{zhongy2009@lzu.cn}}

\abstract{ In this paper, we consider a de Sitter thick brane model in a pure
geometric Weyl integrable five-dimensional space-time, which is a generalization of
Riemann geometry and is invariant under a so-called Weyl rescaling. We find a
solution of this model via performing a conformal transformation to map the Weylian
structure into a familiar Riemannian one with a conformal metric. The metric
perturbations of the model are discussed. For gravitational perturbation, we get the
effective modified P$\ddot{\text{o}}$schl-Teller potential in corresponding
Schr$\ddot{\text{o}}$dinger equation for Kaluza-Klein (KK) modes of the graviton.
There is only one bound state, which is a normalizable massless zero mode and
represents a stable 4-dimensional graviton. Furthermore, there exists a mass gap
between the massless mode and continuous KK modes. We also find that the model is
stable under the scalar perturbation in the metric. The correction to the Newtonian
potential on the brane is proportional to $e^{-3 r \beta/2}/r^2$, where $\beta$ is
the de Sitter parameter of the brane. This is very different from the correction
caused by a volcano-like effective potential.}


\begin{document}

\section{Introduction}

 The possibility that our world is trapped in a
four-dimensional submanifold (called brane) embedded in a fundamental
multi-dimensional space-time (called bulk) has been increasing interest during
recent years (see \cite{Rubakov,CsabaCsaki} for summary of introduction, and
reference therein). The strong motivations originate from the string/M theory. With
the help of the brane scenario, one could possibly solve some disturbing problem of
high-energy physics, such as the hierarchy problems (the problem of why the
electroweak scale $M_{EW}\sim 1$ TeV is so different from the Planck scale
$M_{pl}\sim 10^{16}$ TeV ) and the cosmological constant problem \cite{arkanihamed1,
arkanihamed2, Randall1, Randall2, ccp, CGS}. In this scenario, Standard Model matter
fields are confined to a 3-brane, while the gravity could propagate in the whole
bulk. The most well-known models within are Arkani-Hamed-Dimopoulos-Dvali (ADD)
\cite{arkanihamed1, arkanihamed2} and Randall-Sundrum (RS) \cite{Randall1, Randall2}
models put forward in the end of 90's. They can achieve a Newtonian gravitation law
in macroscopic scale, which is compatible with the observational fact. What
preeminence in RS model is that it suggests a possibility that the extra dimensions
may not be compact any more, namely the size of extra dimensions can be even
infinite.

However, both the models above are assumed that the brane is infinitely thin.
Although with the idealized models many interesting results have been obtained, from
a more realistic point of view, a brane should have the thickness. For this reason,
the emphasis of study has shifted toward the thick brane scenario
\cite{dfgk,grem,cehs,campos,wang,arias,Dzhunushaliev0809,Dzhunushaliev0904,Herrera-Aguilar0910,
Almeida0901,lyx2007,lyx2008,lyx2009a,lyx2009b,lyx2009c,CTA,ZY,Bazeia1,Bazeia2,Bazeia3,Bazeia4,FMP,OGP,KL,FM}.
An interesting feature of thick brane scenario based on gravity coupled to scalar is
that one can achieve a brane naturally without introducing a delta function by hand
in the action to create a brane. And the scalar could further provide the
``material" to make the thick brane. In many multidimensional field theories coupled
to gravity there are solutions of topological defects. They have lead to a richer
variety of thick brane worlds \cite{Dzhunushaliev0904}.

Most investigations of brane worlds are considered in Riemann geometry. In this
paper, we are interested in the thick branes based on a Weyl-integrable
geometry, which is a generalization of Riemann geometry.
\begin{eqnarray}
\text{Metric-affine}\xrightarrow{Q-tr(Q)=0} \text{Weyl-Cartan}
\left\{  {\begin{array}{*{20}c}
   {\xrightarrow{T\neq 0,Q=0}\text{Riemann-Cartan}} \hfill \\
   {\xrightarrow{T=0,Q\neq 0}\text{Weyl}}\hfill\\
   {\xrightarrow{T=0,Q=0}\text{Riemann}} \hfill\\
    \end{array} } \right.
\label{Relation_of_various_Geometry}
\end{eqnarray}

As is shown in (\ref{Relation_of_various_Geometry}) \cite{Puetzfeld}, we can
see there are two type generalizations of Riemann geometry, namely, the
Riemann-Cartan and the Weyl geometry. In Riemann-Cartan geometry, the
connection is asymmetric and the antisymmetric piece of the connection is
represented as torsion $T$ introduced by Cartan. In contrast to Riemann-Cartan
geometry, the connection is no longer metric compatible in Weyl geometry, i.e.,
it involves a geometric scalar $\omega$ in the definition of connection and the
covariant derivative of metric tensor is non-vanishing. The field related to
this violation of the metricity condition is called nonmetricity $Q$. So Weyl
geometry allows for possible variations in the length of vectors during
parallel transport. When $T=0$ the Riemann-Cartan geometry degenerates into
Riemann, and when $Q=0$ the Weyl geometry also degenerates into Riemann
geometry. More precisely, a Weyl geometry is an affine manifold specified by
$(g_{MN} , \omega )$, where $g_{MN}$ is the metric tensor and $\omega$ is the
geometric scalar $\omega$. Since a conformal transformation can map a Weyl
manifold into a Riemann one, the particular type of gauge geometries is called
conformally Weyl or Riemann integrable space-time \cite{Barbosa-Cendejas:2006}.
On the other hand, Weyl integrable manifold is invariant under a Weyl rescaling
(see, (\ref{Weyl_rescaling}) for detail), when this invariance is broken, the
Weyl scalar function $\omega$ transforms into an observable field which
generates the smooth thick brane configurations. Thus in Weyl geometry the
fundamental role in the generation of thick brane configurations is ascribed to
the geometric Weyl scalar field, which is not a bulk field now.

In Refs.
\cite{arias,Barbosa-Cendejas:2006,Barbosa-Cendejas:2005,Barbosa-Cendejas:2008},
the authors have studied brane world scenario in the frame of Weyl geometry, and
achieved various solutions respected to Minkowski thick branes. For most of
these branes \cite{Barbosa-Cendejas:2006,Barbosa-Cendejas:2005}, there exist a
single bound state which represents a stable 4-dimensional graviton and the
continuum spectrum of massive modes of KK excitations without mass gap for a
volcano potential. While in \cite{arias}, the KK spectrum is quantized for an
infinite square-like potential wall, and in \cite{Barbosa-Cendejas:2008}, there
exist one massless bound state, one massive KK bound state and the delocalized
continuum spectrum for a modified P$\ddot{\text{o}}$schl-Teller potential.
These give a claim that Weylian structures mimic, classically, quantum behavior
does not constitute a generic feature of these geometric manifolds
\cite{Barbosa-Cendejas:2005}. In Refs. \cite{Liu0803,Liu0708}, the authors have
considered the localization and mass spectrum problems of matter fields on
these various Minkowski thick branes. An interesting study in \cite{Liu0708}
shows that for scalars there are two bound states (one is normalizable massless
mode), for spin one vectors there is only one normalizable massless bound
state, and for spin half fermions, the total number of bound states is
determined by the coupling constant $\eta$: when $\eta=0$, there are no any
localized fermion KK modes including zero modes for both left and right chiral
fermions, and when $\eta>0$ ($\eta<0$), the number of bound states of right
chiral fermions is one less (more) than that of left chiral fermions and in
both cases ($\eta>0$ and $\eta<0$), only one of the zero modes for left chiral
fermions and right chiral fermions is bound and normalizable.

We will consider a de Sitter thick brane based on gravity coupled
to scalars on a Weyl integrable manifold in this paper, and try to
find a solution of de Sitter thick brane from a pure geometrical
Weyl action in five dimensions. The organization of the paper is
as follows: In Sec. \ref{sec2}, we briefly review the Weyl
geometry. In Sec. \ref{sec3}, we consider a de Sitter thick brane
solution on the weyl manifold and find a solution of the model. In
Sec. \ref{sec4}, we briefly consider the gravitational and scalar
perturbations of the structure. In Sec. \ref{sec5}, we derive the effective Newtonian potential on the brane.  And finally, the conclusion is
given in Sec. \ref{secConclusion}.

\section{Weyl geometry}\label{sec2}

 A non-Riemann generalization of Kaluza-Klein theory
is based on the following five-dimensional action
\cite{arias}:
\begin{equation}
S_5^W  = \int_{M_5^W } {\frac{{d^5 x\sqrt {\left| g \right|} }}
{{16\pi G_5 }}e^{ - \frac{3} {2}\omega } \left[ {R - 3\xi \left(
{\nabla \omega } \right)^2  - 6U\left( \omega  \right)} \right]},
\label{Weyl_5D_action}
\end{equation}
where $M_5^W$ is a five-dimensional Weyl-integrable manifold
specified by the pair $(g_{MN},\omega)$, $g_{MN}$ is the metric
$(M,N=0,1,2,3,5)$ and $\omega$ is a scalar function. In this
manifold the Weylian Ricci tensor reads $ R_{MN}  = \Gamma
_{MN,A}^A - \Gamma _{AM,N}^A  + \Gamma _{MN}^P \Gamma _{PQ}^Q  -
\Gamma _{MQ}^P \Gamma _{NP}^Q $, where the affine connection of
$M_5^W$ is $ \Gamma _{MN}^P  = \{ _{MN}^P \}  - \frac{1}
{2}(\omega _{,M} \delta _N^P  + \omega _{,N} \delta _M^P  - g_{MN}
\omega ^{,P} ) $ with $\{ _{MN}^P \}$ the Christoffel symbol. The
parameter $\xi$ is a coupling constant, and $U(\omega)$ is a
self-interaction potential for the scalar field. Since the scalar
field $\omega$ enters in the definition of the affine connection
of the Weyl manifold, the Weyl action is actually pure
geometrical.

Weyl integrable manifold is invariant under the Weyl rescaling,
\begin{equation}
g_{MN}  \to \Omega ^2 g_{MN} ,~~~\omega  \to \omega  + \ln \Omega ^2
,~~~\xi  \to \xi /(1 + \partial _\omega  \ln \Omega ^2 )^2,
\label{Weyl_rescaling}
\end{equation}
where $\Omega^2$ is a smooth function on $M_5^W$. In general, this
invariance is broken by the self-interaction potential $U(\omega)$.
But from the relation (\ref{Weyl_rescaling}), we know that  $U
\rightarrow \Omega^{-2}U$  is the only transformation preserves such
an invariance. Thus the potential will be $U(\omega)=\lambda
e^{-\omega}$, where $\lambda$ is a coupling constant. When the Weyl
invariance is broken, the scalar field could transform from a
geometric object into an observable degree of freedom which
generates the smooth thick brane configurations.

In order to find solutions of the above theory, we shall use the
conformal technique to map the Weyl frame into the Riemann one.
Under the conformal transformation $g_{MN}=e^\omega \hat g_{MN}$,
the Weylian affine connection becomes the Christoffel symbol
$\Gamma ^P_{MN}\rightarrow\{ _{MN}^P \}$, and the Weylian Ricci
tensor becomes the Riemannian Ricci tensor. In consequence, via
the conformal transformation, one recovers the Riemannian
structure on the manifold $M_5^R$. Now the Weylian action
(\ref{Weyl_5D_action}) is mapped into a Riemann one:
\begin{equation}
S_5^R  = \int_{M_5^R } {\frac{{d^5 x\sqrt {\left| {\hat g} \right|}
}} {{16\pi G_5 }}\left[ {\hat R - 3\xi (\hat \nabla \omega )^2  -
6\hat U(\omega )} \right]}, \label{Weyl_map_into_Riemann_5D_action}
\end{equation}
where $\hat U(\omega)=e^\omega U(\omega)$ and all the hatted
magnitudes and operators refer to the Riemann frame. Thus, one
obtains a theory which describes five-dimensional gravity coupled to
a scalar field with a self-interaction potential in the Riemann
frame.

By considering the variation of the metric $\hat{g}^{MN}$ and the
scalar function $\omega$, we have the set of field equations from
(\ref{Weyl_map_into_Riemann_5D_action}):
\begin{subequations}\label{Weyl_mapinto_R_equations}
\begin{eqnarray}
  \hat G_{MN}  &=&  - \frac{3}{2}\xi \hat g_{MN} (\hat \nabla \omega )^2
       + 3\xi \hat \nabla _M \omega \hat \nabla _N \omega
       - 3\hat g_{MN} \hat U(\omega ) ,  \label{Weyl_mapinto_R_equations_a} \\
  \hat \square \omega  &=& \frac{1}{\xi }
       \frac{{d\hat U(\omega )}} {{d\omega }} .
      \label{Weyl_mapinto_R_equations_b}
\end{eqnarray}\end{subequations}

\section{The model}\label{sec3}

 In Refs.
\cite{arias,Barbosa-Cendejas:2006,Barbosa-Cendejas:2005,Barbosa-Cendejas:2008},
the authors studied thick brane solutions to the theory
(\ref{Weyl_5D_action}), and they considered the metric that respects
four-dimensional Poincar$\acute{\text{e}}$ invariance:
\begin{equation}
ds^2_5=e^{2A(y)} \eta_{\mu \nu} dx^\mu dx^\nu + dy^2,
\end{equation}
i.e., the Minkowski thick brane solutions in Weyl Geometry, where
$e^{2A(y)}$ is the warp factor and $y$ stands for the extra
coordinate. The localization and mass spectrum problems of matter
fields on the Minkowski thick branes were discussed in Refs.
\cite{Liu0803,Liu0708}.

The authors  of \cite{arias} have considered the problem of
$Z_2$-symmetric manifolds. They have chosen $\xi = -(1 + k)/(4k) $
and left $k$ as an arbitrary parameter except for $k = -4/3$. The
following solution was found:
\begin{subequations}\label{Solution1}
\begin{eqnarray}
    \omega(y)&=& \frac{2k}{4-3k}  \ln \left[\cosh\left(\sqrt{\frac{4-3k}{1-k}2\lambda}~y\right) \right], \label{Solution1a} \\
    e^{2A(y)} &=& \left[\cosh\left(\sqrt{\frac{4-3k}{1-k}2\lambda}~y\right) \right]^{\frac{2}{4-3k}}.\label{Solution1b}
\end{eqnarray}\end{subequations}

The authors of \cite{Barbosa-Cendejas:2006} have considered another simplified case
when $k = -4/3 $ and left $\xi$ as an arbitrary parameter except for $\xi = -1/16$.
Their solution is
\begin{subequations}\label{Solution2}
\begin{eqnarray}
    \omega(y) &=& -\frac{2}{1+16\xi}\ln\left\{\frac{\sqrt{-8\lambda
    (1+16\xi)}}{c_1}\cosh \left[c_1 \left(y-c_2 \right)\right] \right\}, \label{Solution2a}\\
    e^{2A(y)} &=& \left\{\frac{\sqrt{-8\lambda
    (1+16\xi)}}{c_1}\cosh \left[c_1 \left(y-c_2 \right)\right] \right\}^{\frac{3}{2(1+16\xi)}}.\label{Solution2b}
\end{eqnarray}\end{subequations}

Obviously, the first solution (\ref{Solution1}) is symmetric with
respect to $y$ coordinate, and the second solution is also a
symmetric one if we perform a translation of $y$ axis.

In this paper, we consider a de Sitter thick brane solution in a
pure geometric five-dimensional Weyl space-time. The line element
for the space-time with planar-parallel symmetry is assumed as
\begin{eqnarray}
 ds_5^2 &=& g_{MN}dx^M dx^N\nonumber \\
        &=&  e^{2A(y)} q_{\mu \nu} dx^\mu dx^\nu + dy^2 \nonumber \\
        &=& e^{2A(y)} \left( - dt^2  + e^{2\beta t} dx^i dx^i \right)
            + dy^2, \label{Weyl_dS_metric}
\end{eqnarray}
where $\beta$ is the de Sitter parameter and related to the 4-dimensional
cosmological constant of the brane. Under the conformal transformation
$g_{MN}=e^\omega \hat g_{MN}$, the de Sitter metric (\ref{Weyl_dS_metric}) is mapped
into
\begin{eqnarray}
      d\hat{s}_5^2
        &=&\hat{g}_{MN}dx^M dx^N  \nonumber \\
        &=&  e^{2\sigma (y)} \left( - dt^2  + e^{2\beta t} dx^i dx^i \right)
            + e^{ - \omega (y)} dy^2 ,
\label{Weyl_mapinto_R_dS_metric}
\end{eqnarray}
where $2\sigma =2A-\omega$.

Now the expressions for Ricci tensor and the scalar curvature in the
Riemann frame read
\begin{eqnarray}
  \begin{array}{l}
      \hat R_{00}  = \left(\sigma '' + 4\sigma '^2  + \frac{1}
      {2}\sigma '\omega ' \right)e^{2A}  - 3\beta ^2 ,  \\
      \hat R_{ij}  =  - \left(\sigma '' + 4\sigma '^2
          + \frac{1}{2}\sigma '\omega '\right)
          e^{2\left( {A + \beta t} \right)} \eta _{ij}
          + 3\beta ^2 e^{2\beta t} \eta _{ij} , \\
      \hat R_{55}  =  - 4\left(\sigma '' + \sigma '^2  + \frac{1}
      {2}\sigma '\omega '\right),  \\
      \hat R =  - 4\left(2\sigma '' + 5\sigma '^2 + \sigma '\omega
      ' \right)e^\omega + 12\beta ^2 e^{ - 2\sigma },
  \end{array}
\label{Weyl_map_into_Riemann_Ricci}
\end{eqnarray}
where the prime denotes derivative with respect to the fifth
coordinate $y$. The five-dimensional stress-energy tensor is given
by
\begin{equation}
\hat{T}_{MN}=\hat{R}_{MN}-\frac{1}{2}\hat{g}_{MN}\hat{R},
\end{equation}
thus, its four-dimensional and five-dimensional components are
given through the following expressions:
\begin{eqnarray}
    \hat T_{00}  &=&  - 3\left( {\sigma '' + 2\sigma '^2  + \frac{1}
    {2}\sigma '\omega '} \right)e^{2A}  + 3\beta ^2  , \nonumber\\
    \hat T_{ij}  &=& 3\left( {\sigma '' + 2\sigma '^2  + \frac{1}
        {2}\sigma '\omega '} \right)e^{2\left( {A + \beta t} \right)} \eta _{ij}
        - 3\beta ^2 e^{2\beta t} \eta _{ij}  , \\
    \hat T_{55}  &=& 6\sigma '^2  - 6\beta ^2 e^{ - 2A} . \nonumber
\label{Weyl_map_into_Riemann_T}
\end{eqnarray}

For simplicity, using the pair of variables $X\equiv \omega'$ and
$Y\equiv 2A'$, we obtain a set of equations from
(\ref{Weyl_mapinto_R_equations}),
 (\ref{Weyl_mapinto_R_dS_metric}) and
(\ref{Weyl_map_into_Riemann_Ricci}):
\begin{subequations}\label{FieldEqs}
\begin{eqnarray}
      &&X' + 2XY - \frac{3} {2}X^2  = \frac{1} {\xi }\frac{{d\hat U(\omega
      )}} {{d\omega }}e^{ - \omega } ,  \\
      &&Y' + 2Y^2  - \frac{3} {2}XY =
      \left(\frac{1} {\xi }\frac{{d\hat U}} {{d\omega }} - 4\hat U \right)e^{ - \omega
      }  + 6\beta ^2 e^{ - 2A}, \\
      &&Y^2 - 2XY + \left(1 - \xi \right)X^2  =
      -2e^{-\omega}  \hat U(\omega ) + 4\beta ^2 e^{ - 2A}.
\end{eqnarray}\end{subequations}

Though we have got the field equations by performing a conformal
transformation $\hat{g}_{AB}  = e^{ - \omega } g_{AB}$ to map the
Weyl manifold structure into the Riemann one, the solutions
corresponding to the Eqs. (\ref{FieldEqs}) are harder to get than
the similar equations in Ref. \cite{arias} for the flat brane
(just set $\beta=0$ in Eqs. (\ref{FieldEqs})). So, in order to
simplify the model, we make a coordinate transformation of the form
$dy =e^{A(z)}dz$. Then the metric (\ref{Weyl_mapinto_R_dS_metric})
will be rewrote as:
\begin{eqnarray}
      d\hat{s}_5^2= e^{2\sigma (z)} \left(q_{\mu \nu}dx^\mu dx^\nu +dz^2 \right)= e^{2\sigma (z)} \left( - dt^2
      + e^{2\beta t} dx^i dx^i + dz^2 \right).
\label{Weyl_mapito_R_C_metric}
\end{eqnarray}

Now repeating our calculation above, the expressions for the Ricci
tensor and the scalar curvature corresponding to the metric
(\ref{Weyl_mapito_R_C_metric}) in the Riemann frame read:
\begin{eqnarray}
   \begin{array}{l}
         \hat R_{00}  = \sigma '' + 3\sigma '^2  - 3\beta ^2   , \\
         \hat R_{ij}  =  - \left(\sigma '' + 3\sigma '^2  - 3\beta ^2 \right)e^{2\beta t} \eta _{ij}, \\
         \hat R_{55}  =  - 4\sigma '' , \\
         \hat R =  - \left(8\sigma '' + 12\sigma '^2  - 12\beta ^2 \right)e^{ - 2\sigma } . \\
   \end{array}
\end{eqnarray}
The stress-energy tensor is given through the following
expressions:
\begin{eqnarray}
      \hat T_{00}  &=&  - 3\left( {\sigma '' + \sigma '^2  - \beta ^2 } \right),  \nonumber \\
      \hat T_{ij}  &=& 3\left( {\sigma '' + \sigma '^2
      - \beta ^2 } \right)e^{2\beta t} \eta _{ij}  , \label{S-E_Tensors} \\
      \hat T_{55}  &=& 6\left(\sigma '^2  - \beta ^2 \right).\nonumber
\end{eqnarray}

So the set of field equations with the new conformally metric
(\ref{Weyl_mapito_R_C_metric}) now read
\begin{subequations}\label{FieldEqs2}
\begin{eqnarray}
      \xi \omega '^2   &=&  \sigma '^2  - \sigma '' - \beta ^2  , \\
      \hat U(\omega )&=& \frac{1}
      {2}e^{ - 2\sigma } \left(-3\sigma '^2  - \sigma '' + 3\beta ^2 \right) , \\
      \frac{1}{\xi}\frac{{d\hat U(\omega )}}
      {{d\omega }} &=& e^{ - 2\sigma } \left(3\sigma '\omega '
      +\omega''\right),
\end{eqnarray}\end{subequations}
where the prime denotes derivative with respect to $z$. If we choose
the parameter $\xi =1/3$, and define $\hat V (\omega)=3\hat U
(\omega)$, we can rewrite Eq. (\ref{FieldEqs2}) as follows:
\begin{subequations}\label{FieldEqs3}
\begin{eqnarray}
      \omega '^2  &=&3\left(\sigma '^2  - \sigma '' - \beta ^2 \right)  , \\
      \hat V(\omega ) &=& \frac{3}
      {2}e^{ - 2\sigma } \left( - 3\sigma '^2  - \sigma '' + 3\beta ^2 \right),  \\
      \frac{{d\hat V(\omega )}}
      {{d\omega }} &=& e^{ - 2\sigma } \left(3\sigma '\omega ' + \omega ''\right).
\end{eqnarray}\end{subequations}
This set of equations is identical with the field equations of the
de Sitter thick brane in the Riemann geometry
\cite{wang,mns1,mns2,irreg1,irreg2,Goetz,gm,Liu0901}. Here, we can
achieve a $Z_2$-symmetric solution refer to $z$ coordinate when we
consider a symmetric domain wall de Sitter expansion in five
dimensions for a sine-Gordon potential \cite{Koley0407,Liu2008c}
\begin{equation}
 \hat V(\omega ) =3\hat U(\omega )= p\cos^2 (q\omega)+v
\label{Potential_V}
\end{equation}
with $p=15\beta^2$, $q=1/\sqrt{3}$, $v=-15\beta^2/2$ and $\beta>0$. The warp factor of the metric
(\ref{Weyl_mapito_R_C_metric}) on Riemann manifold and the scalar
$\omega$ are
\begin{subequations}\label{SolutionOnRiemannManifold}
\begin{eqnarray}
 e^{2\sigma(z) }  &=& \text{sech} \left(2\beta z\right), \label{solution_warpfactor}\\
 \omega(z)  &=& \frac{\sqrt{3}}{2} \arctan
             \big[\sinh (2{\beta z})\big].\label{solution_omega}
\end{eqnarray}\end{subequations}

\begin{figure}[h]
\begin{center}
\includegraphics[width=7cm]{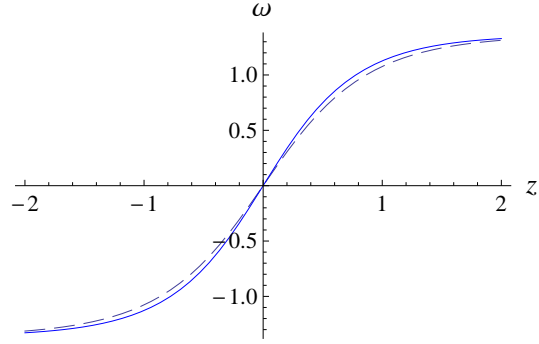}
\end{center}
\caption{The shapes of the scalar $\omega$ in $z$ coordinate. The
parameter $\beta$ is set to $\beta=0.9$ for the dashed line, and
$\beta=1$ for the thin line.} \label{Fig_scalar_z}
\end{figure}

\begin{figure}[htb]
\begin{center}
\includegraphics[width=7cm]{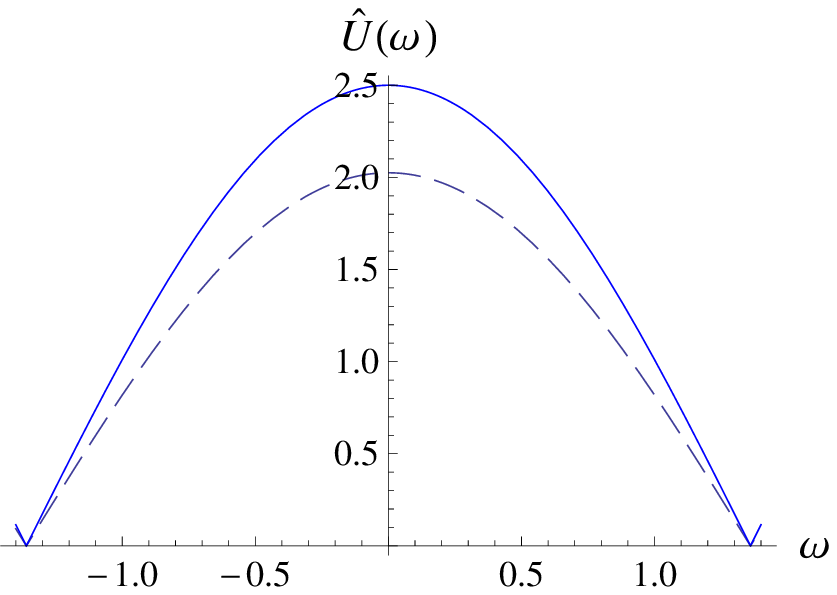}
\includegraphics[width=7cm]{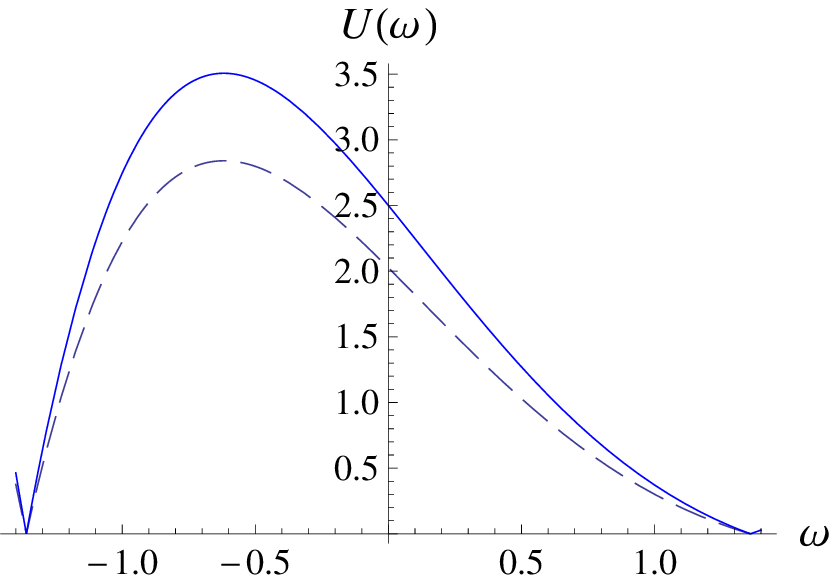}
\end{center}
\caption{The shapes of the potential $\hat{U}(\omega)$ on the Riemann
manifold (left) and the potential $U(\omega)$ on the Weyl manifold
(right). The
parameter $\beta$ is set to $\beta=0.9$ for the dashed line, and
$\beta=1$ for the thin line.}
\label{Fig_potential_z}
\end{figure}
\begin{figure}[htb]
\begin{center}
\includegraphics[width=7cm]{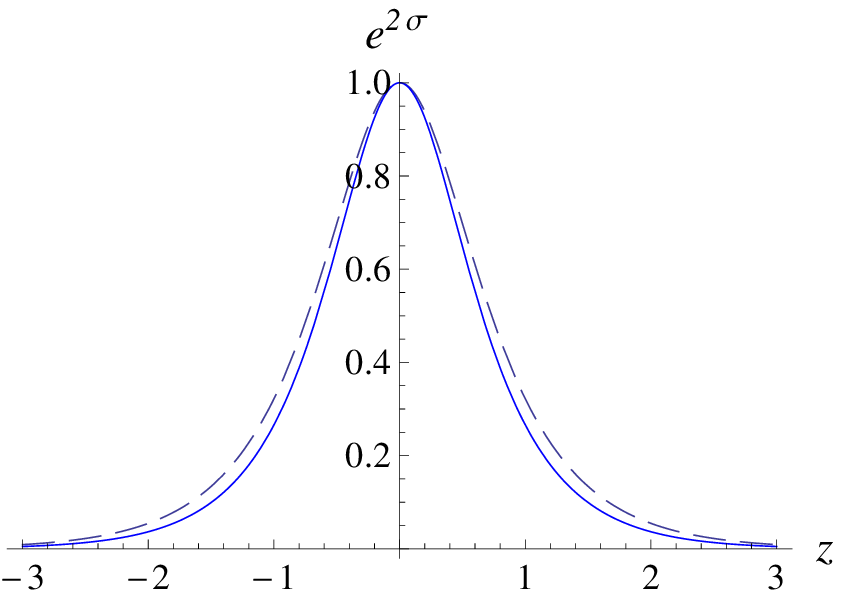}
\includegraphics[width=7cm]{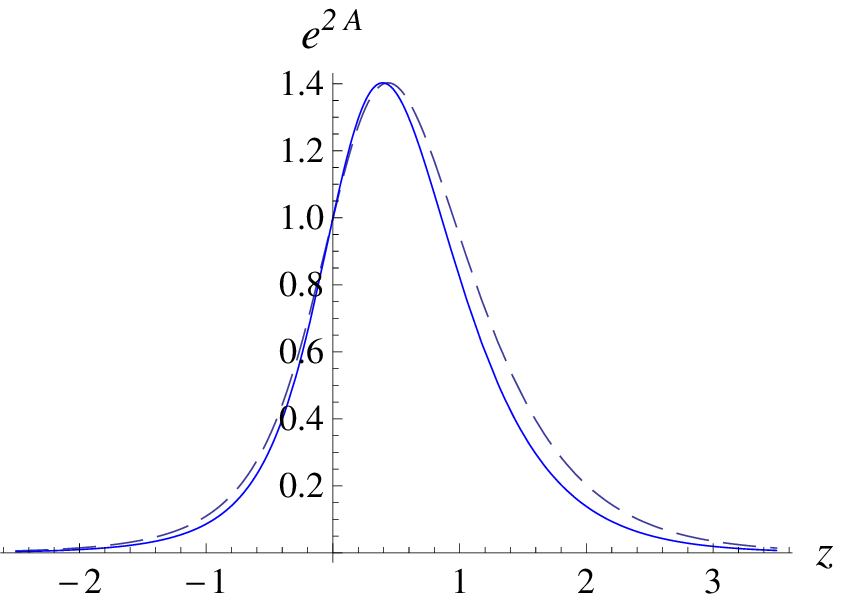}
\end{center}
\caption{The shapes of the warp factor $e^{2\sigma}$ on Riemann manifold (left)
and the warp factor $e^{2A}$ on Weyl manifold (right). The
parameter $\beta$ is set to $\beta=0.9$ for the dashed line, and
$\beta=1$ for the thin line.}
 \label{Fig_warpfactor_z}
\end{figure}

From $\hat U (\omega)=e^\omega U(\omega)$ and Eq.
(\ref{Potential_V}), we have the potential in the original action
(\ref{Weyl_5D_action}) on the Weyl manifold:
\begin{equation}
      U(\omega ) =5\beta^2 \left[  \cos^2 \left(\frac{\omega}{\sqrt{3}}\right)-\frac{1}{2}\right]e^{-\omega}.
\end{equation}
From the relation $e^{2\sigma}=e^{2(A-\omega)}$, the warp
factor $e^{2A(z)}$ is given by
\begin{equation}
     e^{2A(z)}  = e^{\sqrt{3}\arctan[\sinh (2{\beta z})]}
      \text{sech} \left(2\beta z\right).
\end{equation}

The shapes of the scalar $\omega$, the potentials $\hat U (\omega)$
in the Riemannian structure and $U (\omega)$ in the Weylian one, the
warp factors $e^{2\sigma}$ on the Riemann manifold and $e^{2A}$ on
the Weyl manifold are shown in Figs. \ref{Fig_scalar_z},
\ref{Fig_potential_z} and \ref{Fig_warpfactor_z}, respectively. We
can see that the scalar field takes values $\pm \sqrt{3}\pi/4$ at
$z \rightarrow \pm \infty$, corresponding to the two minima of the
potential with cosmological constant $\Lambda=0$. So the scalar is
actually a kink solution, which provides a thick brane realization
of the brane world as a domain wall in the bulk. In the Riemannian
structure, the potential $\hat U (\omega)$ and the warp factor are
symmetric, but in the Weylian one, they are asymmetric.

\begin{figure}[h]
\begin{center}
\includegraphics[width=7cm]{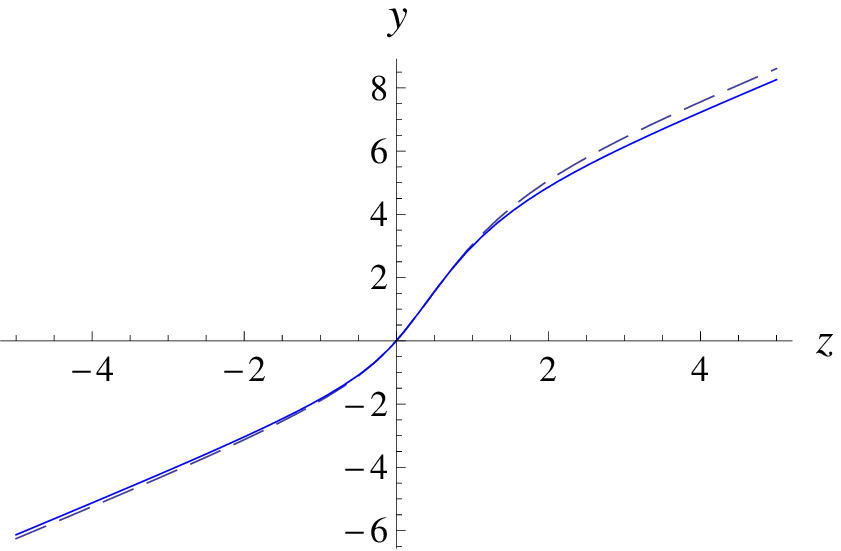}
\includegraphics[width=7cm]{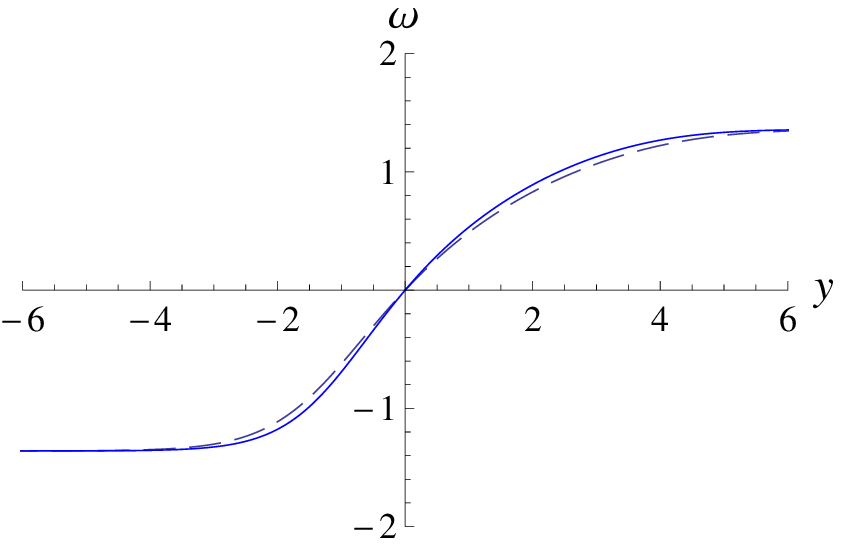}
\end{center}
\caption{The relation of the coordinates $y$ and $z$ (left) and the
scalar field $\omega$ in $y$ coordinate (right). The
parameter $\beta$ is set to $\beta=0.9$ for the dashed line, and
$\beta=1$ for the thin line.}
\label{Fig_relation_y_z}
\end{figure}

Now from $dy=e^{A(z)}dz$, we can calculate the relation between the
original space-time coordinate $y$ in metric
(\ref{Weyl_mapinto_R_dS_metric}) and $z$ in
(\ref{Weyl_mapito_R_C_metric}). However, we could only achieve a
numerical function curve $y=y(z)$ in Fig.  \ref{Fig_relation_y_z}
(left). The scalar field in $y$ coordinate is plotted in Fig.
\ref{Fig_relation_y_z} (right).

\begin{figure}[htb]
\begin{center}
\includegraphics[width=7cm]{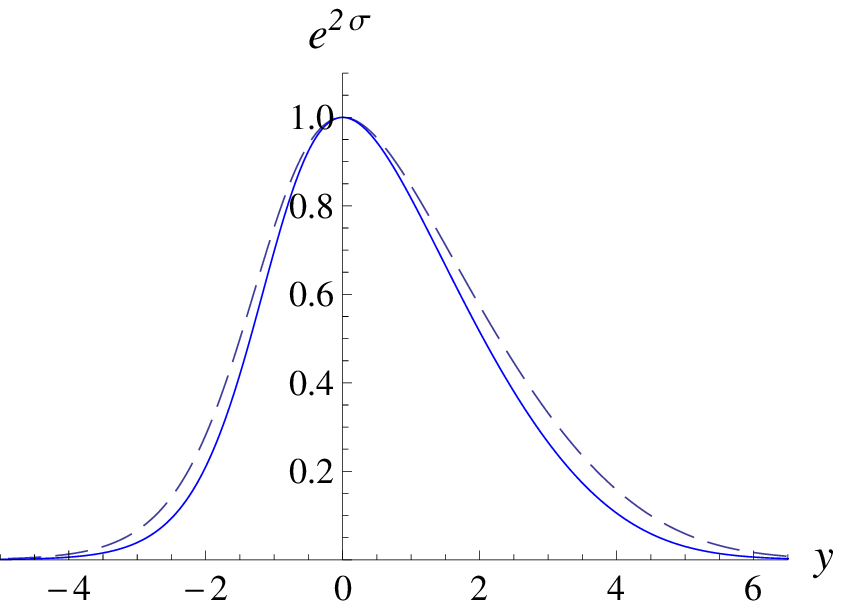}
\includegraphics[width=7cm]{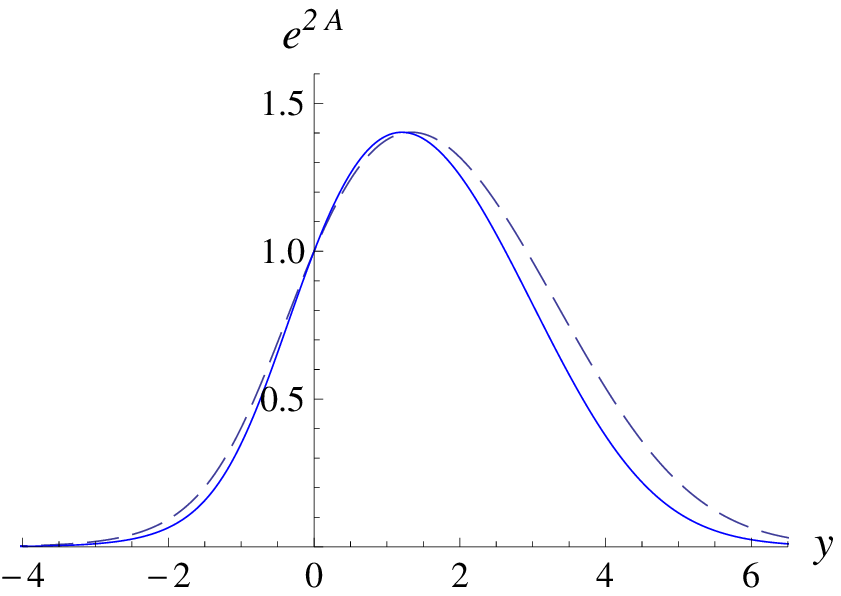}
\end{center}
\caption{The shapes of the warp factors $e^{2\sigma(y)}$ on the
Riemann manifold (left) and $e^{2A(y)}$ on the Weyl manifold
(right). The
parameter $\beta$ is set to $\beta=0.9$ for the dashed line, and
$\beta=1$ for the thin line.}
\label{Fig_warpfactor_y}
\end{figure}

We can find that the scalar $\omega(y)$ is asymmetric but still a
kink-like solution in $y$ coordinate. We display the shapes of the
warp factors $e^{2\sigma (y)}$ and $e^{2A(y)}$ in Fig.
\ref{Fig_warpfactor_y}. The figure shows that the warp factor
$e^{2\sigma (y)}$ is no longer symmetric even in the Riemannian
structure, though the maximum is still at the origin of the $y$
coordinate. So it will not preserve $Z_2$-symmetry in the Riemann
manifold in the $y$ coordinate.

\begin{figure}[htb]
\begin{center}
\includegraphics[width=7cm]{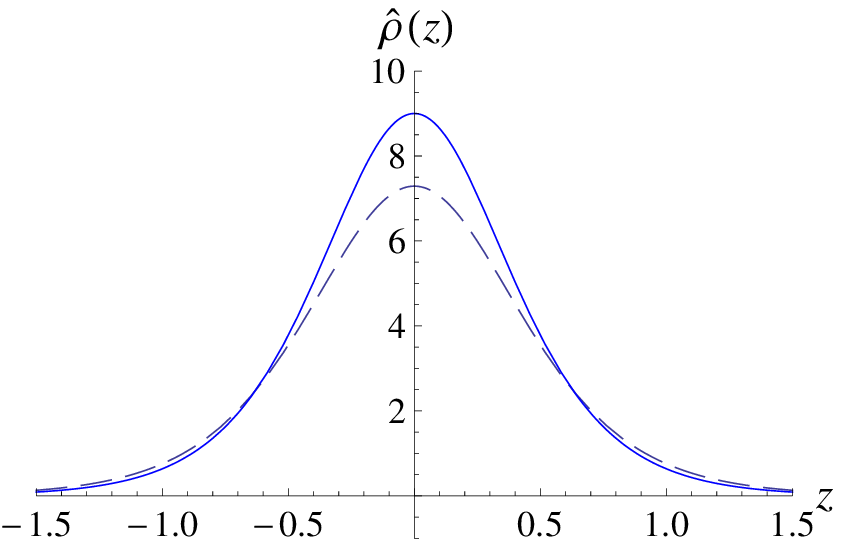}
\includegraphics[width=7cm]{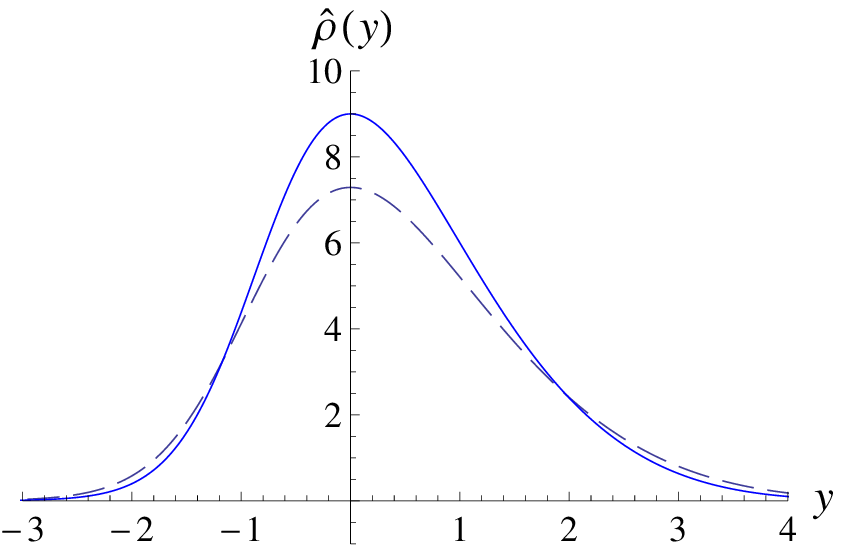}
\end{center}
\caption{The shapes of the energy density $\hat \rho$ in $z$
coordinate (left) and in $y$ coordinate (right) on the Riemann
manifold. The
parameter $\beta$ is set to $\beta=0.9$ for the dashed line, and
$\beta=1$ for the thin line.}
\label{Fig_energydensity}
\end{figure}

The energy density of the scalar matter on the Riemann manifold is
given by the null-null component of the stress-energy tensor
(\ref{S-E_Tensors}):
\begin{eqnarray}
 \hat \rho(z)  &=& \hat T_{00}
   = - 3\left( {\sigma '' + \sigma '^2  - \beta ^2 } \right)  \nonumber \\
  &=& 9\beta ^2\text{sech}^2( 2{\beta z}).
\end{eqnarray}
The shape of the energy density $\hat \rho $ is plotted in Fig.
\ref{Fig_energydensity} in $z$ and $y$ coordinates, respectively. The figures
clearly show that the energy density $\hat \rho$ distributes along extra coordinate,
so it does not stand for a thin brane any more, and $1/\beta$ plays the role of the
brane thickness. The energy density has a maximum at the origin of the coordinate,
and vanishes asymptotically at $y=\pm \infty$, so the scalar matter mainly
distributes on the brane. The energy density is symmetric in $z$ coordinate while
asymmetric in $y$.

\section{Metric perturbations}\label{sec4}

 Firstly, we  consider the gravitational perturbation
in the metric (\ref{Weyl_mapito_R_C_metric}):
\begin{eqnarray}
 d\hat{s}^2_5&=&e^{2\sigma (z)}\left[\left(q_{\mu \nu}
      +h_{\mu \nu}(x,z)\right)dx^\mu dx^\nu
      +dz^2\right] \nonumber \\
     &=&e^{2\sigma (z)}\left[-dt^2 +e^{2\beta t}dx^i dx^i +h_{\mu
      \nu}(x,z)dx^\mu dx^\nu + dz^2\right].
\label{Weyl_mapito_R_CF_Gpertubation_metric}
\end{eqnarray}
Here $ h_{\mu \nu}$ is the tensor perturbation of the metric, and it
satisfies the transverse traceless condition \cite{dfgk,wang,kks},
\begin{eqnarray}
      {h_\mu}^\mu=\nabla^\nu h_{\mu \nu}=0,
\end{eqnarray}
where $\nabla$ denotes the covariant derivative with respect to the
four-dimensional metric $q_{\mu \nu}$. The equation for $h_{\mu
\nu}$ is given by \cite{wang,kks}
\begin{equation}
      h_{\mu \nu}''+3\sigma' h_{\mu \nu}' + \Box h_{\mu \nu} -2\beta^2
      h_{\mu \nu} =0,
\label{Eqs_of_h_mu_nu}
\end{equation}
where $\Box \equiv q^{\mu \nu} \nabla_\mu \nabla _\nu$. Here we can
define the four-dimensional mass of a KK excitation as
\begin{equation}
      \Box h_{\mu \nu} - 2\beta ^2 h_{\mu \nu} = m^2 h_{\mu \nu}.
\label{KK_excitations}
\end{equation}
Furthermore, we can decompose $h_{\mu \nu}$ in the form
\begin{equation}
      h_{\mu \nu}(x,z) = e^{-3\sigma /2 }\varepsilon _{\mu \nu} (x)\Psi(z),
\label{decompsoseh_mu_nu}
\end{equation}
where $\varepsilon_{\mu \nu}$ satisfies the transverse traceless
condition
\begin{eqnarray}
      {\varepsilon_\mu}^\mu =\nabla^\nu \varepsilon_{\mu \nu}=0.
\end{eqnarray}
Now considering (\ref{KK_excitations}) and
(\ref{decompsoseh_mu_nu}), Eq. (\ref{Eqs_of_h_mu_nu}) can be
rewritten as
\begin{equation}
  -\Psi''(z)+\left(\frac{3}{2}\sigma''
    +\frac{9}{4}\sigma'^2\right)\Psi(z)=m^2\Psi(z).
  \label{KK_Schrodinger_Eqs}
\end{equation}
So it turns into a Schr$\ddot{\text{o}}$dinger equation form, and
the effective potential reads
\begin{eqnarray}
  V_g(z)=\frac{3}{2}\sigma''+\frac{9}{4}\sigma'^2
        =\frac{3}{4}\beta^2\left[3-7\text{sech}^2(2{\beta z})\right].
        \label{KK_G_potential}
\end{eqnarray}

The spectrum of the eigenvalue $m^2$ parameterizes the spectrum of
the observed four-dimensional graviton masses. And obviously, there
exists a zero mode which refers to $m=0$. From Eq.
(\ref{KK_Schrodinger_Eqs}) with $m=0$, we can easily get
\begin{equation}
 \Psi_0(z)=c_0 e^{3\sigma(z)/2}
    =c_0 \text{sech}^{3/4}(2{\beta}z). \label{KK_zero_mode}
\end{equation}
It can easily be seen that this zero mode is normalizable, and
$c_0=\sqrt{\frac{2\beta\Gamma({5}/{4})}{\sqrt{\pi}\;\Gamma({3}/{4})}}$ is a
normalization constant. So the 4D gravity can be produced by the zero mode
\cite{Randall2,dfgk,grem,cehs,wang}.
\begin{figure}[yhtb]
\begin{center}
\includegraphics[width=7cm]{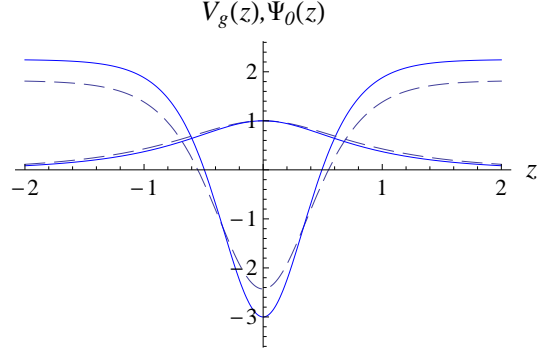}
\end{center}
\caption{The shapes of the potential $V_g(z)$ (concave) and the KK
zero mode $\Psi_0 (z)$ (convex). The
parameter $\beta$ is set to $\beta=0.9$ for the dashed line, and
$\beta=1$ for the thin line.}
 \label{Fig_KK_g_potential}
\end{figure}

The shapes of the potential and the zero mode are plotted in Fig.
\ref{Fig_KK_g_potential}. The figure shows that the gravitational
zero mode localizes around $z=0$, and vanishes at each side
asymptotically. The effective potential is actually a modified
P$\ddot{\text{o}}$schl-Teller potential. From (\ref{KK_G_potential})
we known that the potential barriers at each side approach to a
maximum $V_g^{max}=9\beta ^2 /4$ when $z\rightarrow\pm \infty$, so
there will be exist a set of continuous KK modes $\Psi_m(z)$, and there is a mass
gap between the massless zero mode and the massive continuous KK
modes. The existence of the gap is universal in various kinds of de
Sitter 3-brane model \cite{wang,kks}. It can be shown that there is no any other bound KK mode beside the zero mode for the potential (\ref{KK_G_potential}).

Next, we consider the scalar perturbation. Following the arguments
in Refs. \cite{wang,kks}, one can get the scalar perturbations in the
metric (\ref{Weyl_mapito_R_C_metric}) with the longitudinal gauge:
\begin{eqnarray}
 ds^2_5&=&e^{2\sigma(z)}\left[\left(g_{MN}
        +\delta g_{MN}\right)dx^M dx^N\right] \nonumber \\
     &=&e^{2\sigma(z)}\left[\left(1+2\phi\right)dz^2
        +\left(1+2\psi\right) q_{\mu\nu}dx^\mu dx^\nu\right].
\end{eqnarray}
The equations of the scalar perturbation in the
Schr$\ddot{\text{o}}$dinger-like form are read as \cite{kks}:
\begin{subequations}\label{Relation}
\begin{eqnarray}
&&\delta\omega=\frac{1}{\omega'}\left(-3\psi'+3\sigma'\phi\right), \label{relationA} \\
&&\phi+2\psi=0, \label{relationB}\\
&&\phi(x,z)=\frac{\omega'}{e^{3\sigma/2}}\Phi(x,z),\label{relationC}
\end{eqnarray}
\end{subequations}
\begin{equation}
-\Phi''(x,z)+V_s(z)\Phi(x,z)= \Box \Phi(x,z),
\label{sc_perturbation_eq1}
\end{equation}
where $\Box \equiv q^{\mu \nu} \nabla_\mu \nabla _\nu$, $\delta \omega$ is the perturbation of background scalar $\omega$, and $V_s(z)$
is the effective potential of the scalar perturbation with the
expression
\begin{equation}
 V_s(z)=-\frac{5}{2}\sigma''+\frac{9}{4}\sigma'^2
  +\sigma'\frac{\omega''}{\omega'}-\frac{\omega'''}{\omega'}
  +2\left(\frac{\omega''}{\omega'}\right)^2-6\beta^2.
\end{equation}

Now we decompose $\Phi(x,z)$ according to (\ref{sc_perturbation_eq1}) in the form
\begin{equation}
\Phi(x,z)=X(x)f(z).
\label{Scalardecomposition_}
\end{equation}
Then substitute it into Eq. (\ref{sc_perturbation_eq1}), we get
\begin{subequations}
\begin{eqnarray}
      \Box X(x)&=&m^2 X(x),\label{Eq_of_X}\\
      -f''_m(z)+V_s(z)f_m(z)&=&m^2f_m(z).
      \label{sc_perturbation_eq2}
\end{eqnarray}
\end{subequations}
where $m$ is the 4-dimensional effective mass of the scalar perturbational field. Now Eq. (\ref{Eq_of_X}) can be explicitly expressed as
\begin{eqnarray}
\Box X(x)&&\equiv q^{\mu \nu} \nabla_\mu \nabla_\nu X(x)=q^{\mu \nu}\left(\partial_\mu \partial_\nu-\Gamma^\lambda_{\mu\nu}\partial_\lambda  \right)X(x)\nonumber\\
         &&=\left(-\partial_t^2+e^{-2\beta t}\delta^{ij}\partial_i\partial_j-3\beta \partial_t\right) X(x)=m^2 X(x),
\end{eqnarray}
i.e.,
\begin{eqnarray}
e^{2\beta t}\left(\partial_t^2+3\beta \partial_t+m^2\right)
X(x)=\delta^{ij}\partial_i\partial_j X(x).
\end{eqnarray}
So $X(x)$ can be decomposed as
\begin{equation}
X(x)=T_{km}(t)S_{k}(x^i).
\end{equation}
Then we get
\begin{subequations}
\begin{eqnarray}
    \delta^{ij}\partial_i\partial_j S_k(x^i)=-\vec k^2 S_k(x^i) , \label{Eqs_of_X_xi} \\
     \ddot{T}_{km}(t) + 3 \beta \dot{T}_{km}(t)
      + \left( e^{-2\beta t}{\vec k}^{2} + m^2\right)T_{km}(t)=0.\label{Eqs_of_X_t}
\end{eqnarray}
\end{subequations}
The solution of Eq. (\ref{Eqs_of_X_xi}) is simply achieved as
$S_k(x^i)=ce^{-i\delta_{ij}k^ix^j}$, where $c$ is a constant. Thus we will focus on
Eq. (\ref{Eqs_of_X_t}). The solution is
\begin{eqnarray}
 T_{km}(t)&=&\frac{1}{2\sqrt{2}}\left(\frac{|\vec k|}
      {\beta}e^{-\beta t}\right)^{\frac{3}{2}}
      \bigg[ c_1 ~\Gamma\left(1-\frac{\varsigma}{\beta}\right)
      ~\text{J}\left(-\frac{\varsigma}{\beta},\frac{|\vec k|}
      {\beta}e^{-\beta t}\right) \nonumber \\
   && +c_2 ~\Gamma\left(1+\frac{\varsigma}{\beta}\right)
      ~\text{J}\left(\frac{\varsigma}{\beta},\frac{|\vec k|}
      {\beta}e^{-\beta t}\right) \bigg],
\end{eqnarray}
where $c_1$ and $c_2$ are integration constants,
$\varsigma=\sqrt{\frac{9}{4}\beta^2-m^2}$, and J is the Bessel function.
Now we expand $T_{km}(t)$ about the point $\frac{|\vec k|}{\beta}e^{-\beta
t}\sim 0$ ($t \rightarrow \infty$):
\begin{eqnarray}
 T_{km}(t)&=&c_1~2^{-\frac{3}{2}+\frac{\varsigma}{\beta}}
    \left(\frac{|\vec k|}{\beta}e^{-\beta t}\right)^{\frac{3}{2}
    -\frac{\varsigma}{\beta}}
  + \mathcal{O}\left(\left(\frac{|\vec k|}{\beta}e^{-\beta t}\right)^{\frac{5}{2}
   -\frac{\varsigma}{\beta}}\right)\nonumber\\
 &+&c_2~2^{-\frac{3}{2}-\frac{\varsigma}{\beta}}\left(\frac{|\vec k|}
    {\beta}e^{-\beta t}\right)^{\frac{3}{2}+\frac{\varsigma}{\beta}}
    +\mathcal{O}\left(\left(\frac{|\vec k|}{\beta}e^{-\beta t}\right)
    ^{\frac{5}{2}+\frac{\varsigma}{\beta}}\right) \nonumber \\
 &=&c'_1~e^{-\left(\frac{3}{2}\beta-\varsigma\right)t}
    +\mathcal{O}\left(e^{-\left(\frac{5}{2}\beta-\varsigma\right)t}\right)
    +c'_2~e^{-\left(\frac{3}{2}\beta+\varsigma\right)t}
    +\mathcal{O}\left(e^{-(\frac{5}{2}\beta+\varsigma)t}\right),
\label{Eq_of_gt}
\end{eqnarray}
where $c'_1$ and $c'_2$ are constants independent of time.

\begin{figure}[htb]
\begin{center}
\includegraphics[width=7cm,height=4.7cm]{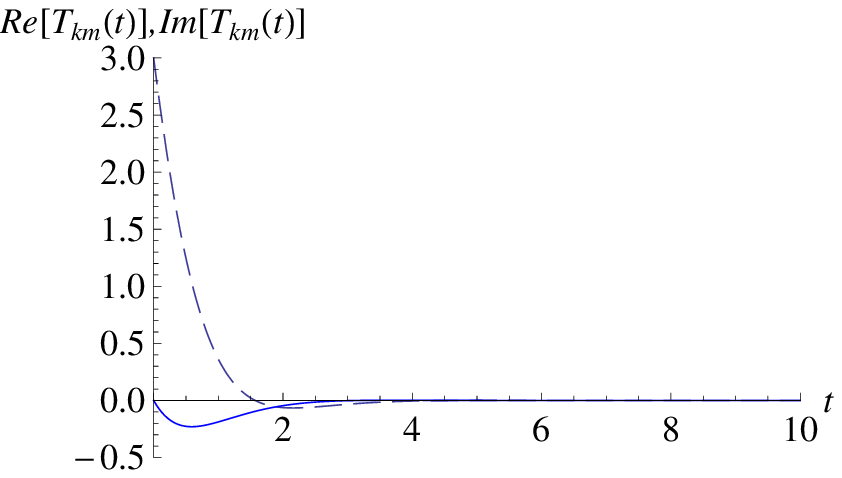}
\includegraphics[width=7cm]{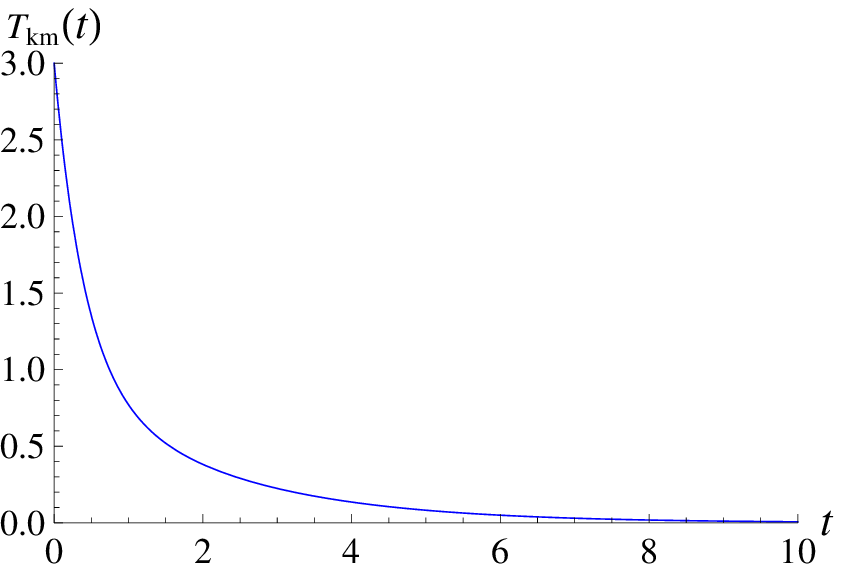}
\end{center}
\caption{The shapes of the function $T_{km}(t)$ in case 1 (left) and
case 2 (right). In case 1, the real part of $T_{km}(t)$ is denoted
by the dash line, and the imaginary part is by the thin line. The
parameters are set to $c'_1=1$, $c'_2=2$, and $\beta=1$.}
\label{Fig_scalar_p_1}
\end{figure}

\textbf{Case 1:}~ When $m^2>\frac{9}{4}\beta^2$, $\varsigma$ is imaginary.
We define $\varsigma = i \zeta$ with $\zeta>0$. Now (\ref{Eq_of_gt}) can
be rewritten as
\begin{equation}
T_{km}(t)= c'_1 e^{-\left(\frac{3}{2}\beta-i \zeta\right)t}+c'_2
e^{-\left(\frac{3}{2}\beta+i \zeta\right)t}. \label{Tcase1}
\end{equation}
Fig. \ref{Fig_scalar_p_1}a shows that, in this case, $T_{km}(t)$
vanishes at $t\rightarrow\infty$, i.e, the scalar perturbation
disappears after a period of time. So we can see the structure is
stable in this case.

\textbf{Case 2:}~ When $0<m^2\leq\frac{9}{4}\beta^2$,
 $\varsigma$ is real and satisfy $0\leq\varsigma<\frac{3}{2}\beta$. Eq. (\ref{Eq_of_gt})
 shows that $T_{km}(t)$ is actually suppressed by $t$,
 and vanishes at $t\rightarrow\infty$, just as shown in Fig.
 \ref{Fig_scalar_p_1}b. So the structure is also
 stable in this case.

\begin{figure}[htb]
\begin{center}
\includegraphics[width=7cm]{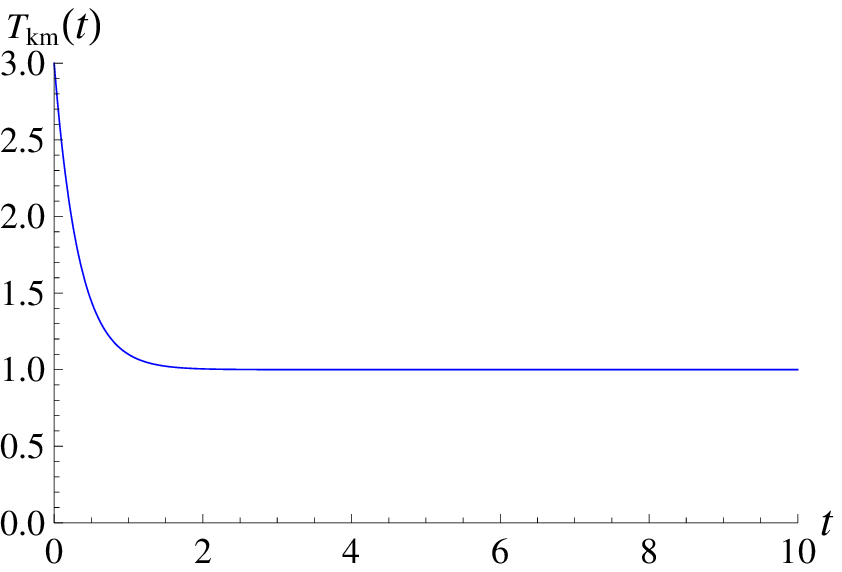}
\includegraphics[width=7cm]{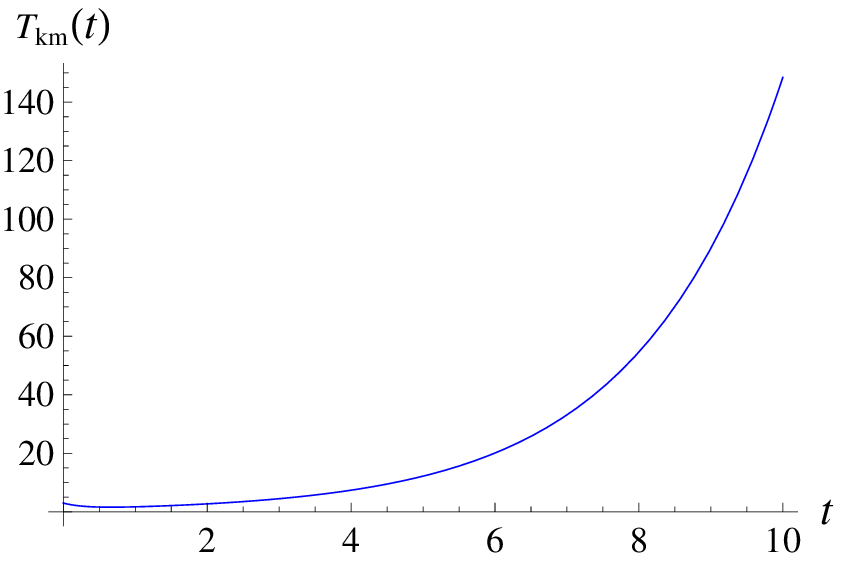}
\end{center}
\caption{The shapes of the function $T_{km}(t)$ in case 3 (left) and
case 4 (right). The parameters are set to $c'_1=1$, $c'_2=2$, and
$\beta=1$.} \label{Fig_scalar_p_2}
\end{figure}

\textbf{Case 3:}~ When $m^2=0$, $\varsigma=\frac{3}{2}\beta$, the high
order terms
$\mathcal{O}\left(e^{-\left(\frac{5}{2}\beta-\varsigma\right)t}\right)$
and
$\mathcal{O}\left(e^{-\left(\frac{5}{2}\beta+\varsigma\right)t}\right)$
in the expanding expression (\ref{Eq_of_gt}) are suppressed by
$t$. While the first term in (\ref{Eq_of_gt}) is a constant
$c'_1=c_1$. If $c_1=0$, $T_{km}(t)$ also converges to zero when
$t\rightarrow\pm\infty$, so the structure is stable. If $c_1\neq
0$, $T_{km}(t)$ will converge to the constant $c_1$ (see Fig.
\ref{Fig_scalar_p_2}a), i.e., some slight scalar perturbations
will always exist but not enormously affect the stability of the
brane. So in this case we can say that the structure is also
stable.

\textbf{Case 4:}~ When $m^2<0$, $\varsigma$ is real and satisfy
$\varsigma>\frac{3}{2}\beta$, the first term in (\ref{Eq_of_gt}) is
divergent at $t\rightarrow\infty$ (see Fig. \ref{Fig_scalar_p_2}b).
This means that in this case the scalar perturbation is divergent,
and the structure is unstable under the scalar perturbation.

From above discussion, it is shown that the structure is stable when
$m^2\geq 0$, while it is unstable when $m^2<0$.

We next focus on Eq. (\ref{sc_perturbation_eq2}), where $V_s(z)$ is
the effective potential of the scalar perturbation. For
(\ref{solution_warpfactor}) and (\ref{solution_omega}), we have
\begin{eqnarray}
 V_s(z)&=&-\frac{5}{2}\sigma''+\frac{9}{4}\sigma'^2+\sigma'
      \frac{\omega''}{\omega'}-\frac{\omega'''}{\omega'}
       +2 \left (\frac{\omega''}{\omega'} \right)^2-6\beta^2 \nonumber \\
  &=&\frac{3}{4}\beta^2\left[3+\text{sech}^2(2 \beta z)\right]. \label{Potential_Vs}
\end{eqnarray}
\begin{figure}[htb]
\begin{center}
\includegraphics[width=7cm]{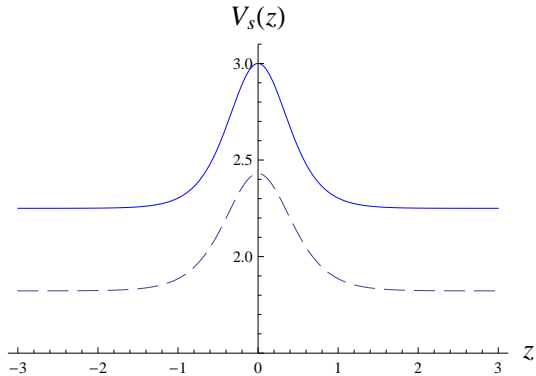}
\end{center}
\caption{The shapes of the potential $V_s(z)$. The
parameter $\beta$ is set to $\beta=0.9$ for the dashed line, and
$\beta=1$ for the thin line.}
 \label{Fig_KK_s_potential}
\end{figure}
From the expression of the above effective scalar perturbational potential, we can
see that $V_s(z)$ approaches to $\frac{9}{4}\beta^2$ when $z\rightarrow \pm \infty$.
Furthermore, $V_s(z)$ satisfies $V_s(z)>\frac{9}{4}\beta^2$ and is always convex and
non-negative. Fig. \ref{Fig_KK_s_potential} shows the potential with virous cases
about the parameter $\beta$. From the Schr$\ddot{\text{o}}$dinger-like equation
(\ref{sc_perturbation_eq2}), we can conclude that there are only KK modes with
eigenvalues $m^2>\frac{9}{4}\beta^2$. This coincides with case 1. So the structure
is stable under the scalar perturbations. The explicit argument about the stable
problem of scalar perturbations could be seen in Ref. \cite{kks}.

The coordinate system corresponding to the four-dimensional metric $q_{\mu\nu}$ in
(\ref{Weyl_dS_metric}) is not a global one. However, even if we use a global
coordinate system, we will find the same results as the previous discussion, namely,
the scalar perturbation will ultimately vanish, approach a constant, and diverge for
the cases of $m^2>0$, $m^2=0$, and $m^2<0$, respectively. So this consequence
results from the property of the four-dimensional background space-time. On the
other hand, Eq. (\ref{Eq_of_X}) is just a Klein-Gordon equation. If the background
is a flat Minkowski space-time, the solution is just a plane wave which never
vanishes and diverges. But now the background is a de Sitter space-time. There is a
exponential expansion factor $e^{2 \beta t}$ before the space components, that means
the space will exponentially expand with time. As a consequence, the symmetries of a
de Sitter space-time are very different from that of a Minkowski one. However, the
space is still homogeneous and isotropic in de Sitter case, so the solution respect
to the Eq. (\ref{Eqs_of_X_xi}) has the same form as in a Minkowski space-time, i.e.,
$S_k(x^i)=ce^{-i\delta_{ij}k^ix^j}$. But the solution of time component will change
and we find that the amplitude will ultimately vanish especially for $m^2>0$ in de
Sitter case. Then from Eqs. (\ref{Tcase1}) and (\ref{Scalardecomposition_}), we can
see that $\Phi(x,z)$ will approach to zero. Therefore, from (\ref{Relation}), we
know that the scalar perturbations $\phi$, $\psi$ and $\delta \omega$ will
ultimately vanish.

In what follows, we will address the question of how the perturbations couple to
matter fields on the brane. However, since the scalar perturbations $\phi$, $\psi$ and $\delta \omega$ ultimately vanish, their interaction with matter fields will decouple
finally. So we only consider the interaction of the tensor perturbations couple to
matter fields on the brane, which will lead to the effective Newtonian potential on
the brane.

\section{The Effective Newtonian Potential}\label{sec5}

In Section \ref{sec4}, we have considered the gravitational perturbation and
achieved a Schr$\ddot{\text{o}}$dinger equation (\ref{KK_Schrodinger_Eqs}). In order
to have localized four-dimensional gravity, we should require that the KK modes of
the Schr$\ddot{\text{o}}$dinger equation (\ref{KK_Schrodinger_Eqs}) don't lead to
unacceptably large corrections to the Newtonian potential in four-dimensional
theory. In the thick brane scenario the matter fields in the four-dimensional theory
on the brane would be smeared over the fifth dimension. For simplicity, like in
Refs. \cite{Rubakov, CsabaCsaki, arkanihamed3, Csaki2000, Bazeia2009}, we consider
the gravitational potential between two point-like sources of mass $M_1$ and $M_2$
located at the origin of the fifth-dimension, i.e., $z=0$. This assumption is
justified in case when the thickness of the brane is small compared with the bulk
curvature. The effective potential between these two particles is given by the
exchange of the massless zero mode and KK massive modes. And the zero mode will
cause a four-dimensional Newtonian interaction potential, the continuum KK modes
will produce the correction to the potential. Thus the effective potential is given
by \cite{Csaki2000}:
\begin{equation}
U(r)= G_N\frac{M_1M_2}{r}+\frac{M_1M_2}{M_*^3}\int_{m_0}^{\infty}{dm \frac{e^{-mr}}{r}|\Psi_m(0)|^2},
\label{Newtonian_potential}
\end{equation}
where $G_N=M_4^{-2}/16\pi$ is the four-dimensional coupling constant, i.e., the Newton's
gravitational constant, $M_*=(16\pi G_5)^{-1/3}$ is the fundamental five-dimensional Planck scale, and
$m_0$ is the minimal eigenvalue at which the continuum KK modes start. In our model,
$m_0$ is not zero but $3\beta/2$ because there is a gap in the mass spectrum.

As in \cite{Randall1}, consider the four-dimensional perturbational metric factors in (\ref{Weyl_mapito_R_C_metric})
\begin{equation}
 d\hat s_5^2=e^{2\sigma(z)}\left[\left(q_{\mu \nu}+h_{\mu \nu}(x)\right)dx^\mu dx^\nu+dz^2\right],
\end{equation}
We decompose the five-dimensional action (\ref{Weyl_map_into_Riemann_5D_action}) into a four dimensional and higher dimensional parts, i.e., \begin{eqnarray}
 S_{5}^R \supset M_*^3\int{d^5x\sqrt{\hat g} \hat R}
       \supset M_*^3\int_{-\infty}^{\infty}{dz e^{3 \sigma(z)} \int d^4x \sqrt{\hat g^{(4)}} \hat R^{(4)}}
       =  M_{4}^2\int{d^4x\sqrt{\hat g^{(4)}} \hat R^{(4)}},
\end{eqnarray}
where $\hat R^{(4)}$ is the four-dimensional Ricci scalar made out of $\hat g^{(4)}_{\mu \nu}=q_{\mu \nu}+h_{\mu \nu}(x)$. So we can read off the relation of the effective four-dimensional Plank scale $M_4$ and the fundamental five-dimensional Planck scale $M_*$,
\begin{eqnarray}
  M_4^2 = M_*^3\int_{-\infty}^{\infty}{dz e^{3 \sigma(z)}}
        = \sqrt{\frac{2}{\pi}} ~ \Gamma^2(\frac{3}{4}) ~ \frac{M_*^3}{\beta}
        \approx 1.2 \frac{M_*^3}{\beta}.
\label{Relation_of_M4_M5}
\end{eqnarray}
We note that since $M_4=M_{pl}$ and the fundamental five-dimensional Planck scale
$M_*$ is not fixed, the order of $M_*$ is decided by the de Sitter parameter
$\beta$. However, an analogue relation can be found in RS2 model
\cite{Rubakov,Randall2}, $M_4^2=M_*^3/k$, where $k$ appearing in the warp factor of
RS metric is a scale of order the Planck scale and $k^{-1}$ represents the anti-de
Sitter radius. Thus $M_*\sim M_{pl}$ in RS2 model. From Eq.
(\ref{Relation_of_M4_M5}), if we chose $\beta \sim M_{pl}$, $M_*$ is of order the
Planck scale in our model.

In order to get the continuum modes, substituting (\ref{KK_G_potential}) into
(\ref{KK_Schrodinger_Eqs}) and introducing a new variable $l=2 \beta z$, we can
reform the Schr$\ddot{\text{o}}$dinger equation as a simple form:
\begin{equation}
-\Psi''(l)-\frac{21}{16}\text{sech}^2(l)\Psi(l)=M^2\Psi(l),
\label{KK_Schrodinger_Eqs_l}
\end{equation}
where the prime denotes derivative with respect to the coordinate $l$, and the new
eigenvalue is $M=\sqrt{\frac{m^2}{4\beta^2}-\frac{9}{16}}$. Since in our model, the
continuum KK modes start at $m\geq 3\beta/2$, so when we just consider these
continuum modes, $M$ will be a nonnegative real number. The solution of this
equation is given by a linear combination of the associated Legendre functions:
\begin{equation}
\Psi_M(l)=C_1~\text{P}(\frac{3}{4}, i M, \tanh(l))+C_2~\text{Q}(\frac{3}{4}, i M, \tanh(l)),
\label{KK_modes_l}
\end{equation}
where $C_1$, $C_2$ are M-dependent parameters. When $l\rightarrow\infty$, each KK
mode wave function will approach to a plane wave:
\begin{equation}
\Psi_M(l)=K_1~e^{iMl}+K_2~e^{-iMl}
\end{equation}
with the parameters $K_1$ and $K_2$ given by
\begin{eqnarray}
&&K_1=\frac{2 C_1 - i \pi C_2 \coth(\pi M)}{2 \Gamma(1 - i M)},\nonumber \\
&&K_2=\frac{C_2}{2C_M [2\sinh^2(M \pi)+i\sinh(2 \pi M)]}, \nonumber
\end{eqnarray}
where $C_M=\frac{4^{i M}}{8\pi^{3/2}}(3 + 8iM + 16M^2) \Gamma(1 + i M)
\Gamma(-\frac{3}{2} - 2 i M)$. Further, we chose the parameters $K_1=K_2=1/2$ to
normalize the plane wave function $\Psi_M(l)$, so the parameters $C_1$ and $C_2$ are
fixed as:
\begin{eqnarray}
C_1&=&\frac{\pi}{2}C_M \left[2\cosh^2(\pi M)
        -i\sinh(2\pi M)\right]+ \frac{1}{2}\Gamma(1 - i M),\nonumber\\
C_2&=&-C_M \left[2\sinh^2(M \pi)+ i \sinh(2 \pi M )\right].\nonumber
\end{eqnarray}

\begin{figure}[htb]
\begin{center}
\includegraphics[width=7cm]{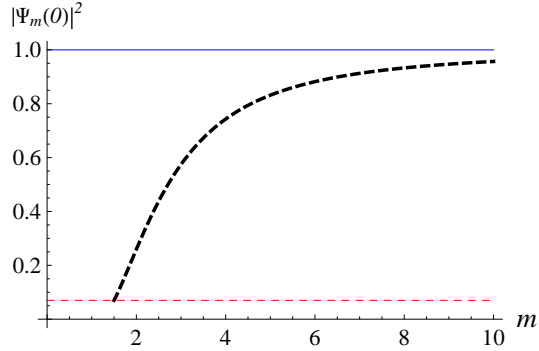}
\end{center}
\caption{The shape of $|\Psi_m(0)|^2$ as a function of $m$.
 The black dashed line represents $|\Psi_m(0)|^2$,
 the red dashed line represents the constant 0.070,
 and the blue thin line represents the constant 1.
 The parameter is set to $\beta=1$.}
\label{kk_modulus_square_at0_with_m}
\end{figure}

Substitute $C_1$ and $C_2$ into Eq. (\ref{KK_modes_l}), we will achieve the whole
wave functions of the continuum KK modes $\Psi_M(l)$. Now replace $l$ with $2 \beta
z$ and $M$ with $\sqrt{\frac{m^2}{4\beta^2}-\frac{9}{16}}$ in the function
$\Psi_M(l)$, we transform $\Psi_M(l)$ back into $\Psi_m(z)$ which refers to the KK
modes wave function in original Schr$\ddot{\text{o}}$dinger equation
(\ref{KK_Schrodinger_Eqs}). We display the curve of $|\Psi_m(0)|^2$ as a function of
$m$ in Fig. \ref{kk_modulus_square_at0_with_m}. It is shown that $|\Psi_m(0)|^2$
will approach to 1 with the increase of the mass $m$. The reason is that the wave
function $\Psi_m(z)$ approximates to a normalized plane wave as the eigenvalue $m$
become large. On the other hand, the value of $|\Psi_{m_0}|^2$ is a constant
$\pi/[\Gamma(1/8) \Gamma(11/8)]^2=0.070$ and independent of the parameter $\beta$.

Now we focus on the Newtonian potential correction term in Eq.
(\ref{Newtonian_potential}). We write the integrand as:
\begin{equation}
I(m)=\frac{e^{-mr}}{r}|\Psi_m(0)|^2.
\label{case0}
\end{equation}
Because it can not be analytically integrated, we will deal with the complex
integrand $I(m)$ under some simple cases for which the integral can easily achieved.

\textbf{Case 1:}~We replace all the KK mode wave functions $\Psi_m(0)$ with the
lowest mode $\Psi_{m_0}(0)$. In this case, $|\Psi_m(0)|^2$ will be replaced by
$\pi/[\Gamma(1/8) \Gamma(11/8)]^2=0.070$. Now the integrant is simply given by
\begin{equation}
I_1(m)=0.070\frac{e^{-mr}}{r}.
\label{case1}
\end{equation}
So in this case the correction of the Newtonian potential is
\begin{equation}
\Delta U_1(r)=0.070\frac{e^{-3 r \beta/2}}{M_*^3}\frac{M_1M_2}{r^2}.
\end{equation}

\textbf{Case 2:}~We replace $|\Psi_m(0)|^2$ with the constant 1, which refers to the limit $|\Psi_{\infty}(0)|^2$ when $m\rightarrow\infty$. In this case, all KK modes are approximately considered as the plane waves. So we have
\begin{equation}
I_2(m)=\frac{e^{-mr}}{r}.
\label{case2}
\end{equation}
The correction of the Newtonian potential is given by
\begin{equation}
\Delta U_2(r)=\frac{e^{-3 r \beta/2}}{M_*^3}\frac{M_1M_2}{r^2}.
\end{equation}

\textbf{Case 3:}~The wave function $|\Psi_m(0)|^2$ approaches to a constant 1 when
$m$ is large and the exponent $e^{-m r}$ is strongly suppressed by large $m$, so the
integrant is mostly determined by the contribution of smaller mass modes. For this
point, we can expand $|\Psi_m(0)|^2$ about the point $m_0=3\beta/2$:
\begin{equation}
|\Psi_m(0)|^2=0.037- 0.079 \frac{m^2}{\beta^2}
  + 0.042 \frac{m^4}{\beta^4}+\mathcal{O}\left(\frac{m^6}{\beta^6}\right).
\end{equation}
So the integrant now can be expressed as
\begin{equation}
I_3(m)=\left(0.037- 0.079 \frac{m^2}{\beta^2}
  + 0.042 \frac{m^4}{\beta^4}\right)\frac{e^{-mr}}{r}. \label{case3}
\end{equation}
In this case, the correction can be write as
\begin{equation}
\Delta U_3(r)=\frac{e^{-3 r \beta/2}M_1M_2}{M_*^3} \left( \frac{0.072}{r^2}  + \frac{ 0.330}{\beta r^3} +
    \frac{0.976}{r^4 \beta^2} +  \frac{1.512}{r^5 \beta^3} +   \frac{1.008}{r^6 \beta^4}\right).
\label{Correction_case3}
\end{equation}

\begin{figure}[htb]
\begin{center}
\includegraphics[width=7cm]{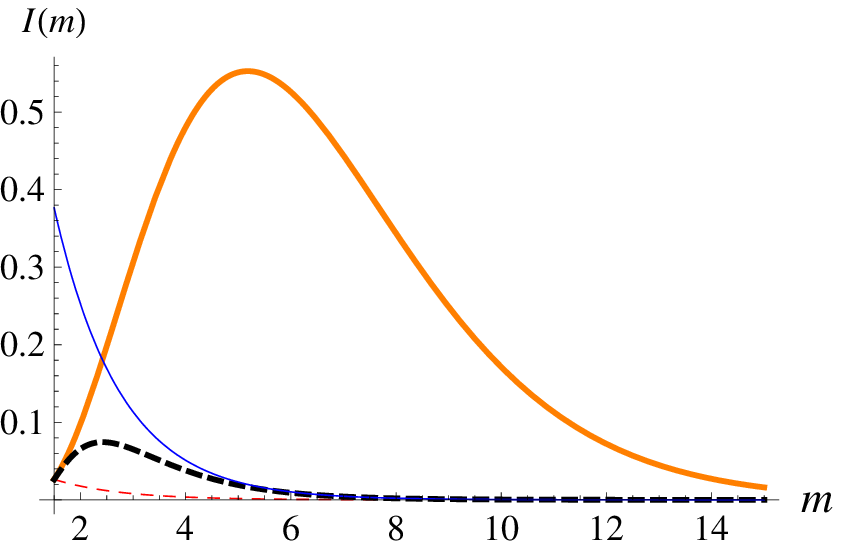}
\includegraphics[width=7cm]{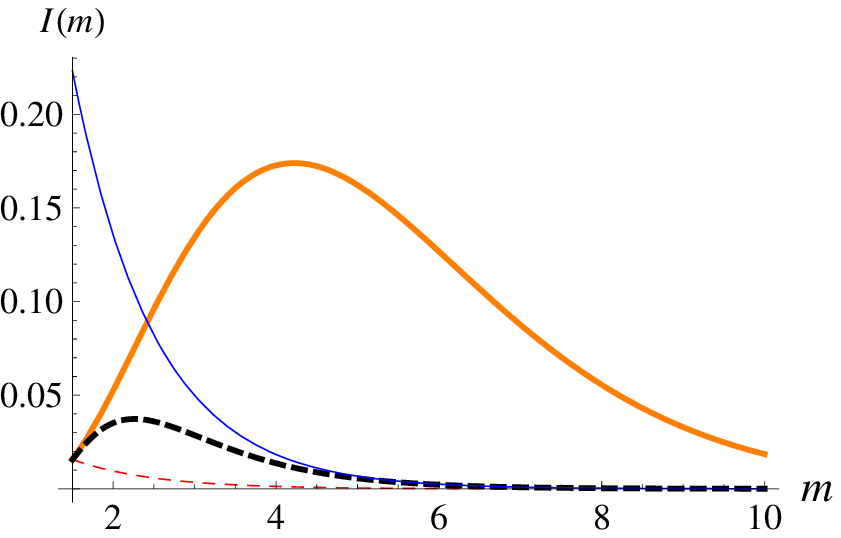}
\end{center}
\caption{The integrand curves of the precise and three approximate cases with fixed
distances $r=0.8$ (left) and $r=1$ (right). The black dashed thick lines represent the
precise integrand curve, the red dashed thin lines represent case 1, the blue thin lines
represent case 2, and the orange thick lines represent case 3. The parameter is set
to $\beta=1$.}
 \label{r_0.8_r_1}
\end{figure}

\begin{figure}[htb]
\begin{center}
\includegraphics[width=7cm]{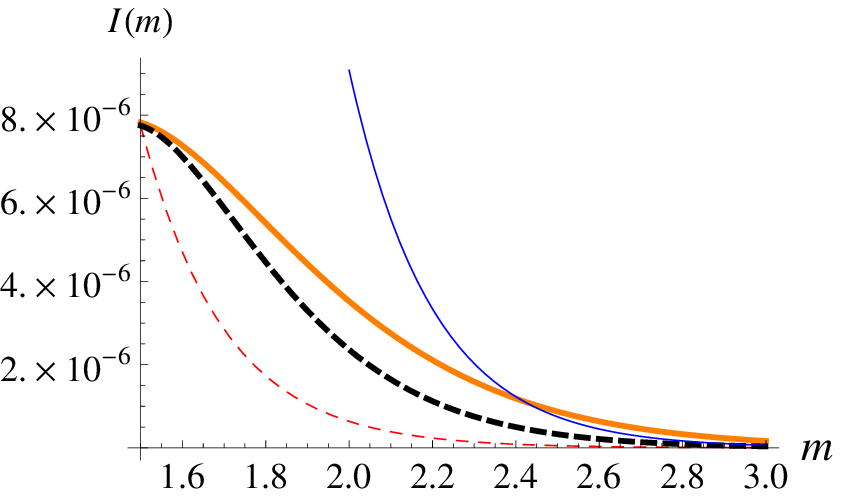}
\includegraphics[width=7cm]{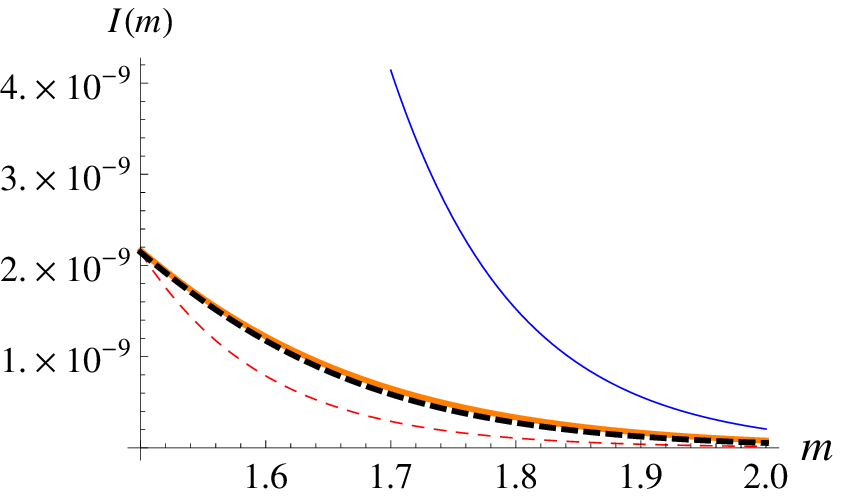}
\end{center}
\caption{The integrand curves of the precise and three approximate cases with fixed
distances $r=5$ (left) and $r=10$ (right). The black dashed thick lines represent the
precise integrand curve, the red dashed thin lines represent case 1, the blue thin lines
represent case 2, and the orange thick lines represent case 3. The parameter is set
to $\beta=1$.} \label{r_5_r_10}
\end{figure}

In order to distinguish which case above is more accurate compared to the original
correction of Newtonian potential, we can compare the precise integrand
(\ref{case0})  with the above three approximate cases (\ref{case1}), (\ref{case2})
and (\ref{case3}) at some fixed distances $r$ of the two particles. Since the area
under an integrand curve represents the integrate value of it, so by comparing the
areas of three cases with the precise condition, we can fixed a suitable form of the
correction. Now we display the curves of the precise and above three approximate
cases with some different fixed distances in Fig. \ref{r_0.8_r_1} and
\ref{r_5_r_10}.


We find that the integrand (\ref{case3}) in case 3 largely deviates from the precise
case when $r$ is small (Fig. \ref{r_0.8_r_1}) but well matches with the precise case
when the distance $r$ is large enough (Fig. \ref{r_5_r_10}). The reason is that we
replace the function $|\Psi_m(0)|^2$ in the integrand (\ref{case0}) with a
polynomial of $m$. The polynomial is a good approximation at small $m$, and even at
large $m$, as long as the distance $r$ is also large. Large $r$ will lead to the
exponent $e^{-mr}$ greatly suppressed, so the polynomial is still a good
approximation. However, when $r$ is small enough, $e^{-mr}$ will not be very little,
but the contribution of the polynomial with large $m$ will be great, in consequence
there will be a obvious derivation with the precise case. This is also can be seen
in Eq. (\ref{Correction_case3}), when $r$ is small enough, the main terms that will
contribute to the correction are the high order terms of $1/r^2$, but when $r$ is
large the main term will be $1/r^2$. So for this consideration we will not use the
correction of case 3 as the approximation of the final correction.  On the other
hand, we can read from the two figures that the precise integrand curve is always
between the curves of case 1 and case 2, and this is obviously, since the Fig.
\ref{kk_modulus_square_at0_with_m} shows that $0.070\leq |\Psi_m(0)|^2 \leq 1$, so
these integrands have the relation $I_1(m)\leq I(m) \leq I_2(m)$. Hence, the precise
correction of Newtonian potential $\Delta U(r)$ must satisfy  $\Delta U_1(r) <
\Delta U(r)< \Delta U_2(r)$, i.e.,
\begin{equation}
0.070\frac{e^{-3 r \beta/2}}{M_*^3}\frac{M_1M_2}{r^2} < \Delta U(r)< \frac{e^{-3 r \beta/2}}{M_*^3}\frac{M_1M_2}{r^2}.
\end{equation}
Though the above discussion, we obtain the suitable correction of Newtonian
potential:
\begin{equation}
\Delta U(r)=\kappa \frac{e^{-3 r \beta/2}}{M_*^3}\frac{M_1M_2}{r^2},
\end{equation}
where the parameter $\kappa$ is a constant satisfied $0.070<\kappa<1$.

So in our mode the effective Newtonian potential can be written as
\begin{equation}
U(r)= G_N\frac{M_1M_2}{r}+\kappa \frac{e^{-3 r \beta/2}}{M_*^3}\frac{M_1M_2}{r^2}.
\end{equation}
This is greatly different from the correction caused by a volcano-like effective
potential \cite{Csaki2000, Bazeia2009}. We can find when the distance $r$ of the two
particles is large, the $1/r^2$ term is a high order term compared with $1/r$, and
the exponent $e^{-3r\beta/2}$ is also greatly suppressed. So in this situation, the
correction is tininess and can be negligible. However, when $r$ is small, $e^{-3 r
\beta/2}/r^2$ will be large and could be the main term of the potential.

\section{Conclusion}\label{secConclusion}

 In this paper, we consider the de Sitter thick brane
solution in Weyl geometry. By performing the conformal
transformation to map the Weyl structure into the familiar Riemann
one, and further, via a coordinate transformation $dy=e^{A(z)}dz$ to
transform the metric into a conformal one, we transform the
structure to a de Sitter thick brane on Riemann manifold. The
solution is
\begin{eqnarray}
      &&U(\omega ) =5\beta^2 \left[  \cos^2 \left(\frac{\omega}{\sqrt{3}}\right)-\frac{1}{2}\right]e^{-\omega}, \\
      &&\omega(z)  = \frac{\sqrt{3}}{2} \arctan \big[\sinh (2{\beta z})\big], \\
      &&e^{2A(z)}  = e^{\sqrt{3}\arctan[\sinh (2{\beta z})]}\text{sech} \left(2\beta z\right).
\end{eqnarray}
The scalar field $\omega$ is a kink solution and preserves symmetry
with respect to $z$ coordinate, while the scalar potential $U(\omega
)$ and warp factor $e^{2A}$ are asymmetric. But actually, $z$
coordinate is not a real physical one. After transforming back to
the physical $y$ coordinate with $dy=e^{A(z)}dz$, we find that all
magnitudes are not preserving symmetry with extra coordinate $y$.
The energy density is still localized around the origin of
coordinate, while our matter distribution on the de Sitter brane is
asymmetric along the extra dimension.

Then we consider the perturbations in the metric (\ref{Weyl_mapito_R_C_metric}) in
Riemann structure. For gravitational perturbation, we get a
P$\ddot{\text{o}}$schl-Teller-like potential for the gravitational KK modes. It is
found that there is a normalizable massless gravitational zero mode localized on the
brane, and exists a mass gap between the zero mode and the massive continuous KK
modes. The existence of such a mass gap also conforms to the universal phenomenon in
various de Sitter 3-brane models. And for scalar perturbation, it is shown that the
scalar perturbations will ultimately vanish, so the solution is stable in this
perturbation.

Finally, we discuss the effective Newtonian potential on the brane, and find the
correction term is proportional to $e^{-3 r \beta/2}/r^2$, which is greatly
different from the correction caused by a volcano-like effective potential. This
result shows that the the perturbations don't lead to unacceptably large corrections
to the Newtonian potential in four-dimensional theory.

\section{Acknowledgments}
We are grateful to the referee, whose comments led to the improvement of this
paper. This work was supported by the Program for New Century Excellent Talents
in University, the National Natural Science Foundation of China (No. 10705013),
the Doctoral Program Foundation of Institutions of Higher Education of China
(No. 20070730055 and No. 20090211110028), the Key Project of Chinese Ministry
of Education (No. 109153), the Natural Science Foundation of Gansu Province,
China (No. 096RJZA055), and the Fundamental Research Funds for the Central
Universities (No. lzujbky-2009-54).

\end{document}